\documentclass{PoS}

\title{Antikaon Interactions with Nucleons and Nuclei - AMADEUS At Da$\Phi$ne}

\ShortTitle{AMADEUS experiment}

\author{\speaker{J. Marton}\\
Stefan-Meyer-Institut f\"ur Subatomare Physik, 1090 Wien, Austria\\
}

\author{K.~Piscicchia\\
        INFN, Laboratori Nazionali di Frascati, 00044 Frascati, Italy\\
Museo Storico della Fisica e Centro Studi e Ricerche Enrico Fermi, Compendio del Viminale, Piazza del Viminale 1, 00184 Roma, Italy\\
E-mail: \email{kristian.piscicchia@lnf.infn.it}
}

\author{C.~Curceanu\\
        INFN, Laboratori Nazionali di Frascati, 00044 Frascati, Italy\\
}

\author{M.~Cargnelli\\
Stefan-Meyer-Institut f\"ur Subatomare Physik, 1090 Wien, Austria\\
}
\author{R.~Del Grande\\
        INFN, Laboratori Nazionali di Frascati, 00044 Frascati, Italy\\
}
\author{L.~Fabbietti\\
        Excellence Cluster 'Origin and Structure of the Universe', 85748 Garching, Germany\\
Physik Department E12, Technische Universit\"at M\"unchen, 85748 Garching, Germany\\
}
\author{G.~Mandaglio\\
        Dipartimento M.I.F.T. dell'Universit\'a di Messina, 98166 Messina, Italy\\
INFN Gruppo collegato di Messina, 98166 Messina, Italy\\
}
\author{M.~Martini\\
        INFN, Laboratori Nazionali di Frascati, 00044 Frascati, Italy\\
Dipartimento di Scienze e Tecnologie applicate, Universit\'a 'Guglielmo Marconi', 00193 Roma, Italy\\
}
\author{P.~Moskal\\
Institute of Physics, Jagiellonian University, 30-059 Krakow, Poland\\
}
\author{A.~Scordo\\
        INFN, Laboratori Nazionali di Frascati, 00044 Frascati, Italy\\
}
\author{D.~Sirghi\\
        INFN, Laboratori Nazionali di Frascati, 00044 Frascati, Italy\\
}
\author{M.~Skurzok\\
 Institute of Physics, Jagiellonian University, 30-059 Krakow, Poland\\
}
\author{I.~Tucakovic\\
        Ru{\dj}er Bo{\u s}kovi{\' c} Institute, Zagreb, Croatia\\
}
\author{O.~Vazquez Doce\\
Excellence Cluster 'Origin and Structure of the Universe', 85748 Garching, Germany\\
Physik Department E12, Technische Universit\"at M\"unchen, 85748 Garching, Germany\\
}
\author{S.~Wycech\\
        National Centre for Nuclear Research, 00681 Warsaw, Poland\\
}
\author{J.~Zmeskal\\
Stefan-Meyer-Institut f\"ur Subatomare Physik, 1090 Wien, Austria\\
}

\abstract{The aim of AMADEUS is to provide unprecedented experimental information on K$^-$ absorption in light nuclear targets, to face major open problems in hadron nuclear physics in the strangeness sector, namely the nature of the $\Lambda$(1405), strongly related to the possible existence of kaonic nuclear clusters, kaons and hyperon scattering cross sections on nucleons and nuclei. These issues are fundamental for a better understanding of the non-perturbative QCD in the strangeness sector. 

AMADEUS step 0 deals with the analysis of the 2004-2005 KLOE collected data. The interactions of the negative kaons produced by the DA$\Phi$NE collider (a unique source of monochromatic low-momentum kaons) with the materials of the KLOE detector, used as active targets, provide samples of K$^-$ absorptions on H, ${}^4$He, ${}^{9}$Be and ${}^{12}$C, both at-rest and in-flight.
A second step deals with the data from the implementation in the central region of the KLOE detector of a pure graphite target, providing a high statistic sample of K$^- \, {}^{12}$C nuclear captures at rest. For the future a new setup, with various dedicated gaseous and solid targets, is under preparation.

}

\FullConference{The 26th International Nuclear Physics Conference\\
		11-16 September, 2016\\
		Adelaide, Australia}

\begin{document}

\section{Introduction}
\label{intro}

The AMADEUS (Anti-kaonic Matter At DA$\Phi$NE: An Experiment with Unraveling Spectroscopy) \cite{amadeus} experiment is conceived to perform a high acceptance and high resolution study of the hadronic interaction processes of kaons with nucleons and light nuclei. The analyses presented in this work refer to the data acquired by the KLOE \cite{kloe} collaboration during the 2004/2005 data taking campaign. The absorption of the low momentum (about 127 MeV/c), almost monochromatic, negatively charged kaons provided by the DA$\Phi$NE factory \cite{dafne}, in the materials of the KLOE detector, supplies high statistic samples of K$^-$ captured on H, $^4$He, $^9$Be and $^{12}$C, both at-rest and in-flight. The aim is to provide experimental constrains for the understanding of the non-perturbative  QCD in the strangeness sector, with important consequences ranging from hadron and nuclear physics to astrophysics.

The $\Lambda(1405)$ is a spin $1/2$, isospin $I=0$ and strangeness $S=-1$ negative parity baryon resonance which decays into $(\Sigma \pi)^0$ through the strong interaction.
Despite the fact that $\Lambda(1405)$ is currently listed as a four-stars resonance in the table of the Particle Data Group (PDG) \cite{PDG}, its nature still remains an open issue.
The three quark picture ($uds$) fails to reproduce the observed properties of this state. A review of the theoretical works, and references to the experimental literature can be found in \cite{Hyodo}. According to the chiral unitary predictions \cite{chiral} the observed shape of the $\Lambda(1405)$ could emerge from the interplay of two poles. A lower mass (about 1380 MeV), broader, pole is mainly coupled to the $\Sigma \pi$ channel, a higher mass (located around 1420 MeV) narrower pole is mainly coupled to the $\bar{\mathrm{K}}$N production channel.
Since the accessible invariant mass, in K$^-$p absorption processes, is influenced by the binding energy of the proton in the hosting nucleus, our strategy is to unveil the presence of the high mass pole by exploiting K$^-$ captures in-flight \cite{piscicchiabormio,scordo}. In this case the kinetic energy of the kaon sets the energy threshold just below the $\bar{\mathrm{K}}$N threshold. The shapes of the $(\Sigma\pi)^0$ spectra are also distorted by the non-resonant production below threshold. A key related issue, which is addressed in the analyses described below, is the investigation of the non-resonant hyperon-pion transition amplitude below threshold.

The strength of the $\bar{\mathrm{K}}$N interaction influences the position of the $\Lambda(1405)$, and the formation of more complex $\bar{\mathrm{K}}$N-multi-nucleon clusters.
For the di-baryonic kaonic bound state ppK$^-$ theoretical
predictions deliver a wide range of binding energies and widths
\cite{plb6}, while the experimental results are contradictory \cite{plb7,plb8,plb9,plb10,mas11,mas12,mas13,masa}.
Moreover, the extraction of the ppK$^-$ signal in K$^-$ absorption experiments is strongly affected by the yield and the shape of the competing K$^-$ multi-nucleon absorption processes.

In Section \ref{sec-1} the features of the DA$\Phi$NE accelerator and the KLOE detector are summarized and the two data samples which are presently under analyses are described. The particle identification procedure is summarised in Section \ref{pid}. Sections \ref{sec-2} and \ref{hyp-pion} are presenting the obtained results and ongoing analyses about the K$^-$ multi-nucleon absorption processes, the search for a ppK$^-$ signal and the investigation of resonant and non-resonant hyperon-pion production in light nuclei. The paper ends with conclusions and perspectives.

\section{The KLOE detector at DA$\Phi$NE, data samples}
\label{sec-1}

DA$\Phi$NE (Double Annular $\Phi$-factory for Nice Experiments) is a double ring $e^+e^-$
collider, designed to work at the center of mass energy of the $\phi$(1020) particle;
$\phi$ meson decays into low momentum ($\simeq\,$ 127 MeV/c) charged kaons, which allows to either stop them, or to
explore the products of their low-energy nuclear absorptions.

The KLOE detector is centered around the electron-positron
interaction region of DA$\Phi$NE and its acceptance amounts to 98\%; 
it consists of a large cylindrical Drift Chamber (DC) \cite{kloedc} and a fine sampling calorimeter consisting of  lead and scintillating fibers \cite{kloeemc}, all immersed in the axially symmetrical magnetic field with a strength of 0.52 T provided by a superconducting solenoid.
The chamber is characterized by excellent position and momentum resolutions. 
Tracks are reconstructed with a resolution in the transverse $R-\phi$ plane
$\sigma_{R\phi}\sim200\,\mathrm{\mu m}$ and a resolution along the $z$-axis $\sigma_z\sim2\,\mathrm{mm}$.
The transverse momentum resolution for low momentum tracks $(50<p<300) \mathrm{MeV/c}$)
is $\frac{\sigma_{p_T}}{p_T}\sim0.4\%$.
The calorimeter is composed of a cylindrical barrel and two endcaps,
providing a solid angle coverage of 98\%.
The volume ratio (lead/fibers/glue=42:48:10) is optimized for high light yield and high efficiency for photons in the range
(20-300) MeV/c. The photon detection efficiency is 99$\%$ for energies larger than 80 MeV and it falls down to 80$\%$ at 20 MeV due to the cutoff introduced by the ADC and TDC thresholds. The position of the clusters along the fibers can be reconstructed with a resolution $\sigma_{\parallel} \sim 1.4\, \mathrm{cm}/\sqrt{E(\mathrm{GeV})}$. The resolution in the orthogonal direction is  $\sigma_{\perp} \sim 1.3\, \mathrm{cm}$. The energy and time resolutions for photon clusters are given by $\frac{\sigma_E}{E_\gamma}= \frac{0.057}{\sqrt{E_\gamma (\mathrm{GeV})}}$ and 
$\sigma_t= \frac{57 \, \mathrm{ps}}{\sqrt{E_\gamma (\mathrm{GeV})}} \oplus 100 \,\, \mathrm{ps}$.

 The DC entrance wall composition is 750 $\mu$m of carbon fiber and 150 $\mu$m
of aluminum foil. Dedicated GEANT Monte Carlo simulations of the KLOE apparatus show that out of the total number of kaons interacting in the DC entrance wall, 
about 81\% are absorbed in the carbon fiber component and the remaining 19\% in the aluminum foil. 
The KLOE DC is filled with a mixture of helium and isobutane (90\% in volume $^4$He and 10\% in volume C$_4$H$_{10}$).

\begin{figure}[htb]
\centerline{%
\includegraphics[width=11.5cm]{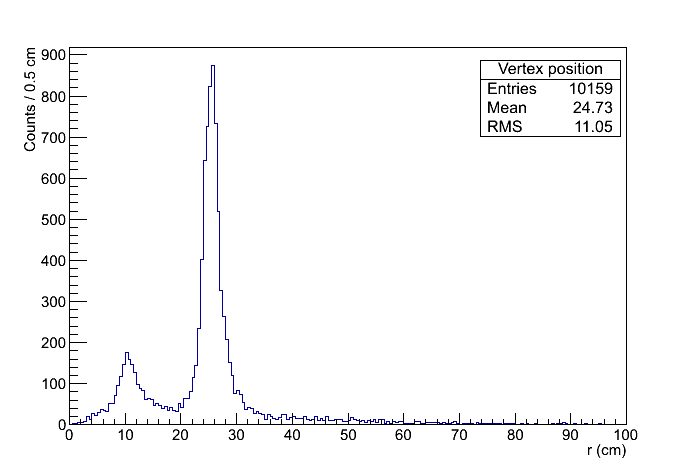}}
\caption{\em Radial position distribution $\rho_\Lambda$, of the the $\Lambda$ decay vertex, for 2004-2005 KLOE data.}
\label{kloe2004}
\end{figure}

Two data samples are presently under analyses. One corresponds to the $\sim$ 1.74 fb$^{-1}$ data collected by the KLOE collaboration during the 2004/2005 data taking, for which the $dE/dx$ information of the reconstructed tracks is available ($dE/dx$ represents the truncated mean of the ADC collected counts due to the ionization in the DC gas).The 
hadronic interactions of negative kaons with the materials of the apparatus are analysed.
The topology of these data is shown in figure \ref{kloe2004}, representing the radial position ($\rho_\Lambda$) of the $\Lambda(1116)$ decay vertex (see Section \ref{pid}). Four components are visible, from inside to outside  we observe $K^-$ absorptions in the DA$\Phi$NE beryllium sphere ($\sim$ 5 cm), the DA$\Phi$NE aluminated beryllium pipe ($\sim$ 10 cm), the KLOE DC entrance wall (aluminated carbon fiber $\sim$ 25 cm) and the long tail originating from $K^-$ interactions in the gas filling the KLOE DC (25-200 cm).
This sample contains rich experimental information on the K$^-$ interactions with the nuclear targets, both at-rest and in-flight \cite{piscicchiabormio}.

The second data sample deals with runs in which a carbon target was used  to increase the statistics. The high purity carbon target (graphite) was realized in summer 2012 and installed inside the KLOE DC, between the beam pipe and the DC entrance wall.
We took data from 6 November to 14 December 2012, for a total integrated luminosity of $\sim$90 pb$^{-1}$, out of which 37 pb$^{-1}$ of reconstructed data were analysed.

The analyses presented in Sections \ref{pid}, \ref{sec-2} and \ref{hyp-pion} refer to the 2004/2005 data sample.

\section{Particle Identification}\label{pid}

The interactions of negatively charged kaons in nuclear matter were investigated by using correlated pairs of hyperon-pion or hyperon-nucleon/nucleus, following the K$^-$ absorptions in H, ${}^4$He, ${}^9$Be and ${}^{12}$C.
%The investigation of the $K^-$ multi-nucleon absorptions and the properties of possible antikaon multi-nucleon bound states proceeds through the analyses of the $\Lambda/\Sigma-p,d,t$, correlations; this last channel is, in particular, extremely promising for the search and characterisation in different nuclear targets of the extremely rare four nucleon absorption process. 
%The search for the $\Lambda(1405)$ is performed through its decay into $\Sigma^0\pi^0$ (purely isospin I=0) and $\Sigma^+\pi^-$.
%Given the excellent momentum resolution for charged particles, also the analysis of the $\Lambda \pi^-$ (isospin I=1) production, both from direct formation process and from internal conversion of a primary produced $\Sigma$ hyperon ($\Sigma \, N \rightarrow \Lambda \, N'$) is presently ongoing. The aim is to measure, for the first time, the module of the non-resonant transition amplitude (compared with the resonant $\Sigma^{*-}$) below threshold. 
The $\Lambda(1116)$ identification proceeds through the reconstruction of the $\Lambda \rightarrow$ p + $\pi^-$ (BR = 63.9 $\pm 0.5 \%$) decay vertex. A
spatial resolution below 1 mm is achieved for vertices found inside the DC volume (evaluated with
Monte Carlo simulations).
The obtained $M_{\mathrm{p}\pi^-}$ invariant mass mean value is 1115.753$\pm$0.002 MeV/c\textsuperscript2 (only statistical error is given, the systematics being under investigation),
with a resolution of $\sigma $=0.5 MeV/c\textsuperscript2. 
The particle identification takes advantage of both $dE/dx$ information from the DC wires and the measurement of the energy released in the Calorimeter, as described in \cite{amadeus}.
%A common vertex between the
%$\Lambda$ candidate and an additional proton or deuteron or triton track is then searched for. 
%As an example, the obtained resolution on the
%radial coordinate ($\rho_{\Lambda p}$) for the $\Lambda p$ vertex is 12 mm, and this topological 
%variable is used to select the $K^-$ absorption processes inside
%the DC wall.
$\Sigma$ particles are identified through their decay into
$\Lambda \gamma$ or p$\pi$ as reported in \cite{piscicchiabormio,scordo}.  
%The photon selection is carried out via its identification in the EMC. Photon candidates are selected by applying a cut on the difference between
%the EMC time measurement and the expected time of arrival of
%the photon within $-1.2 < t < 1.8$ ns.
%Using the time and energy information from the electromagnetic calorimeter, $\pi^0$ particles can be reconstructed through their $\gamma\gamma$ decay.
The K$^-$ absorption vertex position, obtained using the correlated production of the hyperon together with an additional particle (pion, proton etc.) is then used to select the target. As an example, the obtained resolution on the
radial coordinate ($\rho_{\Lambda \mathrm{p}}$) for the $\Lambda$p vertex is 1.2 mm.
Cuts on the absorption vertex radial position were optimised, based on MC simulations and a study of the $\Lambda$ decay path, in order to select the targets with a minimal contamination from other components. More details on the particle identification procedure can be found in \cite{had21}

\begin{figure}[ht]
\centering
\includegraphics[width=15.5cm,clip]{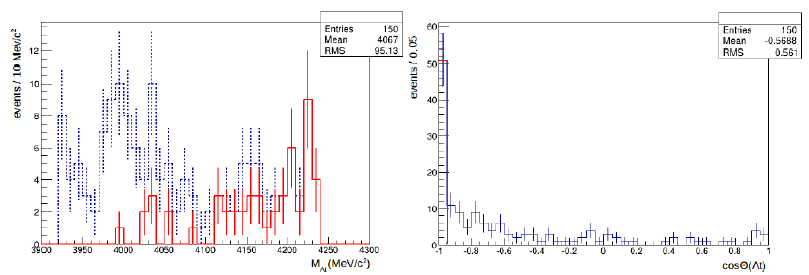}
\caption{\em (Colour online.) $\Lambda t$ invariant mass (left) and $\cos\theta_{\Lambda t}$  (right). The events corresponding to the $cos\theta_{\Lambda t}<-0.95$ selection are shown in red.}
\label{fig-2}       % Give a unique label
\end{figure}

\begin{figure}[ht]
\centering
\includegraphics[width=9cm,clip]{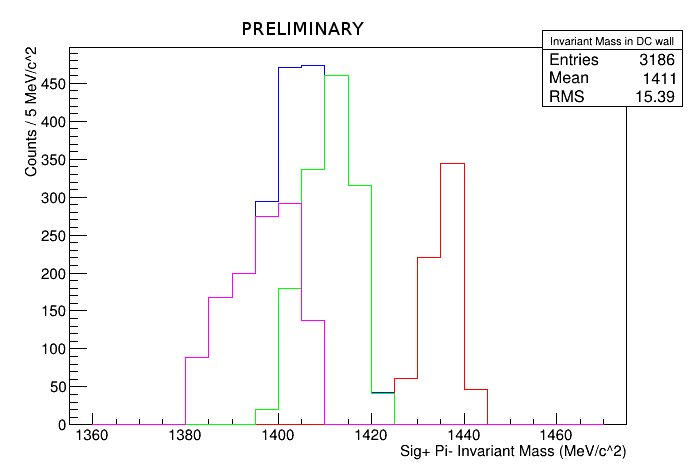}
\caption{\em (Colour online.) $m_{\Sigma \pi}$ invariant mass distributions in-flight (green) and at-rest (violet) in ${}^{12}$C. Blue histogram represents the sum of green and violet histograms. The red distribution refers to K$^-$ absorptions on Hydrogen}
\label{fig-3}       % Give a unique label
\end{figure}

\section{K$^-$ multi-nucleon absorption processes and search for the ppK$^-$ bound state}
\label{sec-2}

In \cite{had21} a high purity sample of $\Sigma^0$p events from K$^-$ captures in the ${}^{12}$C was reconstructed. $\Sigma^0$p is, together with $\Lambda$p, an expected decay channel of the ppK$^-$ cluster, with the advantage to be free from the $\Sigma$ N $\rightarrow \Lambda$ N' conversion processes. The conversions strongly affect the uncorrelated $\Lambda$p production, thus distorting the observed spectra.

A simultaneous fit of the: $\Sigma^0$p invariant mass, the relative angle between
the $\Sigma^0$ and proton in the laboratory system $\cos(\theta_{\Sigma^0 \mathrm{p}})$, the $\Sigma^0$ and
the proton momenta was performed by including the following (simulated) processes:

\begin{itemize}
\item K$^-$A$ \rightarrow \Sigma^0 - (\pi) $p$_{\mathrm{spec}} $(A') (1NA),
\item K$^-$pp$ \rightarrow \Sigma^0$p (2NA),
\item K$^-$ppn$ \rightarrow \Sigma^0 $pn (3NA), 
\item K$^-$ppnn$ \rightarrow \Sigma^0 $pnn (4NA),
\end{itemize}
where 1NA stands for one nucleon absorption, 2NA, 3NA and 4NA for the absorption on 2, 3 and 4 nucleons respectively. Also the Final State Interactions (FSI) of the $\Sigma^0$ and p emerging from a K$^-$pp capture were taken into account.

The yield of the 2NA, when the produced $\Sigma^0$ and p particles are free form any FSI process (we will refer to this component as 2NA-QF), was measured for the first time, with good precision. The obtained results are summarized in Table 1. A second fit was carried out including a ppK$^-$ component, decaying into $\Sigma^0 $p. A systematic scan of possible binding energies and widths, varying within 15-75 MeV and 30-70 MeV respectively, was performed. The best fit resulted in a binding energy of 45 MeV and a width of 30 MeV. The resulting yield, normalised to the number
of stopped K$^-$, is ppK$^-$/K$^-_{\mathrm{stop}}=(0.044\pm0.009$stat$_{-0.005}^{+0.004}$syst$)\times$10$^{-2}$.
The significance of the bound state with respect to a statistical fluctuation was checked by means of an F-test and was found to be significant at the level of 1$\sigma$ only. Although the measured spectra are compatible with the hypothesis of a ppK$^-$ contribution, the significance of the result is not sufficient to claim the discovery of the state. We refer to \cite{had21} for the details of the analysis.

\begin{table}[htbp]
\centering
\begin{tabular}{cccc}
\hline
Process & yield / K$^-_{\mathrm{stop}}\times$10$^{-2}$ & $\sigma_{\mathrm{stat}}\times10^{-2}$ & $\sigma_{\mathrm{syst}}\times10^{-2}$ \\
\hline
2NA-QF     & $0.127$ & $\pm0.019$ & $^{+0.004}_{-0.008}$ \\
2NA-FSI    & $0.272$ & $\pm0.028$ & $^{+0.022}_{-0.023}$ \\
Tot 2NA    & $0.399$ & $\pm0.033$ & $^{+0.023}_{-0.032}$ \\
3NA        & $0.274$ & $\pm0.069$ & $^{+0.044}_{-0.021}$ \\
Tot 3 body & $0.546$ & $\pm0.074$ & $^{+0.048}_{-0.033}$ \\
4NA + bkg. & $0.773$ & $\pm0.053$ & $^{+0.025}_{-0.076}$ \\
\hline
\end{tabular}       % Give a unique label
\caption{Production probability of the $\Sigma^0$p final state for
different intermediate processes normalised to the number of
stopped K$^-$ in the DC wall. The statistical and systematic
errors are shown as well \cite{had21}.}
\end{table}

The measurement of the, extremely rare, 4NA absorption process (K$^-$ + $^4$He$\, \rightarrow \Lambda $t) is presently ongoing. To this aim K$^-$ captures in the gas filling the KLOE DC are exploited, with the goal to pin down the $\Lambda$t 4NA production in $^4$He. Three events compatible with the $\Lambda$t kinematics were identified in \cite{lt1} from K$^-$ stopped in liquid helim. Fourty $\Lambda$t events were measured by the FINUDA collaboration \cite{lt2} from K$^-$ absorptions in different solid targets (${}^{6,7}$Li, ${}^{9}$Be). In these works the 4NA contribution was not disentangled from the other competing processes. In our work the highest statistics ever of correlated $\Lambda $t production was evidenced (150 events). The preliminary $\Lambda$t invariant mass and angular correlation distributions are shown in Fig. \ref{fig-2} left and right respectively.

The signature of the K$^-$ 4NA in $^4$He is the production of back-to-back $\Lambda $t pairs, with the highest energy permitted by kinematics. Such events are represented in red in Fig. \ref{fig-2}, and correspond to the cut $\cos\theta_{\Lambda \mathrm{t}}<-0.95$. The fit of the measured spectra is ongoing.

\section{Y$\pi$ resonant and non resonant production and the shape of the $\Lambda(1405)$}\label{hyp-pion}

The position of the $\Lambda(1405)$ state is determined by the strength of the $\bar{\mathrm{K}}$N attractive interaction, thus influencing the possible formation of $\bar{\mathrm{K}}$ multi-N states formation. When extracting the $\Lambda(1405)$ shape from K$^-$ induced reactions in light nuclear targets (see for example \cite{esmaili}) the hyperon-pion spectroscopy is influenced by the energy threshold, imposed by the last nucleon binding energy. The $m_{\Sigma \pi}$ invariant mass threshold is about 1412 MeV and 1416 MeV, for K$^-$ capture at-rest in ${}^4$He and ${}^{12}$C respectively, thus the K$^-$ absorption at-rest is not sensitive to the high mass pole predicted by chiral unitary models. The $\bar{\mathrm{K}}$N sub-threshold region is accessible by exploiting K$^-$N absorptions in-flight. For a mean kaon momentum of 100 MeV/c the $m_{\Sigma \pi}$ threshold is shifted upwards by about 10 MeV. A second bias is represented by the non-resonant K$^-$N$\rightarrow $Y$\pi$ formation, which gives rise to the production of strongly correlated hyperon-pion pairs. The corresponding $m_{Y \pi}$ invariant masses spectra are narrow (of the order of 10 MeV) and peaked below the $\bar{\mathrm{K}}$N threshold. The $\Lambda \pi$ and $\Sigma \pi$ non resonant transition amplitudes, for K$^-$ capture in light nuclear targets, was never measured. The $\Lambda$ and $\pi^-$ kinematic distributions for K$^-$ captures in ${}^4$He, both at-rest and in-flight, were calculated in \cite{piscicchia}. The momentum probability distribution functions of the emerging hyperon pion pairs, following K$^-$n absorptions, are expressed in terms of the K$^-$n transition amplitudes: the isospin $I=1$ S-wave non- resonant amplitude ($f^{\mathrm{nr}}$) and the resonant $I=1$ P-wave amplitude, dominated by the $\Sigma^-(1385)$. Since the resonant amplitude is well known from direct experiments, the measured total momentum distributions can be used to extract the non-resonant $|f^{\mathrm{nr}}|$ amplitude module below the $\bar{\mathrm{K}}$N threshold. The goal of the ongoing analyses is to measure the contributions and the shapes of the non resonant $\Lambda \pi$ and $\Sigma \pi$ productions. The knowledge of the $(\Sigma \pi)^0$ isospin $I=0$ non-resonant transition amplitude will allow to disentangle the resonant $\Lambda(1405)$ shape. Preliminary $\Sigma^+\pi^-$ invariant mass spectra, from K$^-$ captures in the wall of the KLOE DC, that are not background subtracted nor acceptance corrected, are shown in Fig. \ref{fig-3}. The red hystogram refers to K$^-$ absorptions on Hydrogen, green and violet distributions refer to K$^-$ captures in-flight and at-rest in ${}^{12}$C respectively, the blue distribution is the sum of the green and the violet. The red distribution reflects the non-resonant K$^-$H absorption in-flight, which corresponds to a narrow invariant mass shape peaked below $m_K + m_p + \left<p_K^2 \right>/2m_K$, the third term represents the mean kinetic energy of the non-relativistic kaons. A high statistics sample of in-flight  K$^-{}^{12}$C captures can be separated from the corresponding at-rest absorptions, which is peaked at around 1415 MeV. A spectroscopic study of the $(\Sigma \pi)^0$ production in the sub-$\bar{\mathrm{K}}$N threshold region, opened by the low momentum in-flight capture process, will allow to clarify the nature of the high mass $\Lambda(1405)$ pole.

\section{Conclusions and perspectives}
In this work a broad research program of low-energy K$^-$ induced reactions on light nuclear targets is presented. The $\bar{\mathrm{K}}$N interaction in nuclear matter is investigated through the study of the hyperon resonances properties below the $\bar{\mathrm{K}}$N threshold, and the characterization of K$^-$-multi-nucleon captures processes. The latter is found to strongly impact on the K$^-$-multi-nucleon bound state search, as the K$^-$pp (2NA) absorption overlaps with the bound state in the phase space region where it is expected. 

The elastic and inelastic scattering of hyperons with the residual nucleons, in the final state of the K$^-$ absorption process, were taken into account, when searching for the K$^-$pp bound state, or extracting the K$^-$-multi-nucleon absorption yields, free from final state interactions.
The elastic hyperon-nucleon(s) scattering processes are of particular interest for the measurement of the hyperon-nucleon (multi-nucleon) cross sections, for which the available experimental information is extremely scarce. Moreover, the hyperon-nucleon (multi-nucleon) interaction potentials are fundamental inputs in the determination of the equation of state for the neutron stars, whose structure is strongly debated, following the measurement of two neutron stars exceeding 2M$\odot$ \cite{demorest, nice}. Experimental constrains from the hyperon-nucleon scattering processes are mandatory to guide the theory.

Presently a feasibility study \cite{bazzi_nim,bazzi_jou} is ongoing for the realization of a dedicated AMADEUS experimental setup, in order to deepen and extend the low energy anti-kaon nuclei interaction studies and obtain fundamental input for the study of QCD with strangeness and of neutron stars.

\section*{Acknowledgement}
\noindent
\ignorespaces
\normalfont\fontsize{9pt}{10.8pt}\selectfont
We acknowledge the KLOE Collaboration for their support and
for having provided us the data and the tools to perform the analysis presented in this paper.

\end{document}